\documentclass[conference,letterpaper]{IEEEtran}

\usepackage[utf8]{inputenc} 
\usepackage[T1]{fontenc}
\usepackage{ifthen}
\usepackage{cite}
\usepackage[cmex10]{amsmath} 
                             
\usepackage[utf8]{inputenc} 
\usepackage[T1]{fontenc}    
\usepackage{booktabs}       
\usepackage{amsfonts}       
\usepackage{nicefrac}       
\usepackage{microtype}      
\usepackage{amsmath, amssymb}      
\usepackage{enumerate}
\usepackage{relsize}
\usepackage{tikz} 
\usepackage{pgfplots}
\usepackage{pgfplotstable}
\usepackage{filecontents}
\usetikzlibrary{decorations.pathmorphing} 
\usetikzlibrary{automata}
\usepgfplotslibrary{statistics}
\pgfplotsset{compat = newest}
\pgfkeys{/pgf/number format/set thousands separator = }
\definecolor{chocolate1}{rgb}{0.94,0.7,0.5}

\newcommand{\bigomega}{\makebox{\Large\ensuremath{\omega}}}

\newcommand{\R}{\mathbb{R}}
\newcommand{\C}{{\mathbb C}}

\newcommand{\X}{\bf{X}}
\newcommand{\x}{{\bf x}}

\newcounter{examplecntr}
{\begin{trivlist}\small\item[]\refstepcounter{examplecntr}%
 {\bfseries Example~\theexamplecntr%
  \ifthenelse{\equal{#1}{}}{}{ (#1)}.
}}%
{\end{trivlist}}

\newcounter{propositioncntr}
\newenvironment{proposition}[1][]%
{\begin{trivlist}\item[]\refstepcounter{propositioncntr}%
{\bfseries Proposition~\thepropositioncntr%
  \ifthenelse{\equal{#1}{}}{}{ (#1)}.
}}%
{\end{trivlist}}

\newcounter{remarkcntr}
{\begin{trivlist}\item[]\refstepcounter{remarkcntr}%
{\bfseries Remark~\theremarkcntr%
  \ifthenelse{\equal{#1}{}}{}{ (#1)}.
}}%
{\end{trivlist}}

\newcounter{theoremcntr}
{\begin{trivlist}\item[]\refstepcounter{theoremcntr}%
{\bfseries Theorem~\thetheoremcntr%
  \ifthenelse{\equal{#1}{}}{}{ (#1)}.
}}%
{\hfill$\Box$\end{trivlist}}

\newcommand{\IT}{IEEE Trans.\ Inf.\ Theory}

\begin{document}
\title{Duality for Continuous  Graphical Models} 


\author{%
  Mehdi Molkaraie\\
   Department of Statistical Sciences\\
  University of Toronto\\
\tt{mehdi.molkaraie@alumni.ethz.ch}}


\maketitle

\begin{abstract}
The dual normal factor graph and the factor graph duality theorem have been considered for discrete 
graphical models. In this paper, we show an application of 
the factor graph duality 
theorem to continuous graphical models.
Specifically, we propose a method to solve exactly the Gaussian graphical models defined on the ladder graph 
if certain conditions on the local covariance matrices are satisfied. Unlike the conventional approaches,
the efficiency of the method depends on the position of the zeros in the local covariance matrices. The method and 
details of the dualization 
are illustrated on two toy examples.

\end{abstract}

\section{Introduction}

Graphical models represent the decomposition of multivariate functions
into the product of several local factors. Usually, each local factor depends on a small
subset of the variables. 

The normal factor graph duality theorem has been previously applied to computational problems in discrete graphical 
models (e.g., discrete spin systems and codes on graphs) \cite{MoLo:ISIT2013, AY:2014, Forney:11, MoGo:18}.
In this paper, we consider an application of the duality theorem to continuous graphical models. Our focus is on 
Gaussian graphical models.
These models are extremely useful in many different areas, including spatial statistics \cite{besag1974spatial}
and gene expression studies \cite{dobra2004sparse}. Another reason for the significance of Gaussian graphical models 
is due to the multivariate central limit theorem~\cite[Chapter 3]{anderson1958introduction}.
For more details on Gaussian graphical 
models, see~\cite{dempster1972covariance}, \cite{rue2005gaussian}, \cite[Chapter 5]{lauritzen1996graphical}.

A zero-mean real random vector ${\X}_{N\times 1}$ has an $N$-variate\linebreak Gaussian distribution if it has the following PDF
\begin{equation}
\label{eqn:GaussModel}
p(\x) =  \frac{1}{\textrm{det}(2\pi \mathsf{\Sigma})^{1/2}}
\textrm{exp}\big(-\frac{1}{2}
\x^\intercal
\,\mathsf{\Sigma}^{-1} 
\x\big), \quad \x \in \R^N
\end{equation}
where 
the symmetric positive-definite matrix $\mathsf{\Sigma}^{-1} \in \R^{N\times N}$ is the 
precision (information) matrix and $\mathsf{\Sigma}$ is the corresponding covariance
matrix. We will use the notation $\mathsf{\Sigma} \succ 0$ to indicate that $\mathsf{\Sigma}$ is positive-definite, 
and sometimes denote the sequence $(x_1, x_2, \ldots, x_N)$ by $\x_1^N$.

The structure of a Gaussian graphical model is specified by the precision matrix. Indeed, the
nonzero off-diagonal entries of the precision matrix indicate the presence of an edge that connects
the two corresponding random variables in the graphical model. 
Moreover, according to the pairwise Markov property, two non-adjacent random variables are conditionally independent 
given all the other variables in the model~\cite{lauritzen1996graphical}.

We consider the problem of solving exactly 
Gaussian graphical models for a specific model (i.e., the ladder graph) if certain 
conditions on the local covariance and the local precision matrices are satisfied. By ``exactly solve" 
we mean that the method can efficiently compute the normalization 
constant in~(\ref{eqn:GaussModel}), which boils down to an efficient computation 
of $\textrm{det}(\mathsf{\Sigma})$. 
In our framework, we consider the dual normal factor graph (NFG) of the Gaussian ladder graph. Our approach relies on 
this key property of
multivariate Gaussian distributions: in the Fourier transform of a
multivariate Gaussian distribution the precision matrix is replaced by the covariance 
matrix in the exponent~\cite[Chapter 2]{anderson1958introduction}.

Contrary to the standard approaches, our method relies on the position of the zeros of the covariance 
matrices associated with the local factors, where the local covariance matrices are required to 
have cycle-free graphical representations. 

In order to perform exact inference, we first transform the Gaussian
graphical model into a cycle-free NFG via the Fourier transformation of the 
local factors. In cycle-free (Gaussian) graphical models, exact 
inference can then be done efficiently via the (Gaussian) belief propagation algorithm~\cite{weiss2001correctness}.

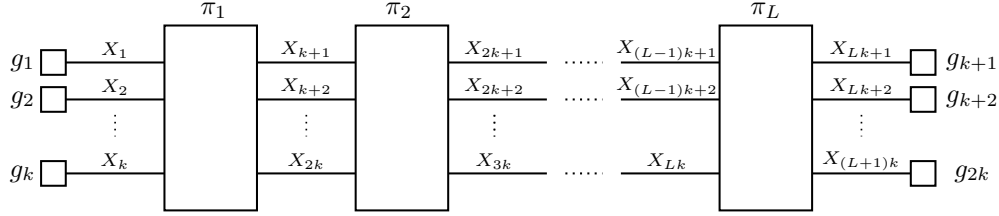
\begin{figure*}[t]
  \centering
  \begin{tikzpicture}[scale=0.82]
 \draw [line width=0.27mm] (0, 0) rectangle (1.5,3.0);
 \draw [line width=0.27mm] (3.1, 0) rectangle (4.6,3.0);
\draw [line width=0.27mm] (9, 0) rectangle (10.5,3.0);
\draw [line width=0.27mm] (-2.0, 2.2) rectangle (-1.6,2.6);
\draw [line width=0.27mm] (-2.0, 1.6) rectangle (-1.6,2.0);
\draw [line width=0.27mm] (-2.0, 0.4) rectangle (-1.6,0.8);
\draw [line width=0.27mm] (12.1, 2.2) rectangle (12.5,2.6);
\draw [line width=0.27mm] (12.1, 1.6) rectangle (12.5,2.0);
\draw [line width=0.27mm] (12.1, 0.4) rectangle (12.5,0.8);
 \draw (-0.8,2.6) node{${\scriptstyle X_1}$};
 \draw (-0.8,2.0) node{${\scriptstyle X_2}$};
 \draw [line width=0.22mm, dotted] (-0.8,1.55) -- (-0.8,1.2);
 \draw (-0.8,0.8) node{${\scriptstyle X_k}$};
 \draw (2.3,2.6) node{${\scriptstyle X_{k+1}}$};
\draw (2.3, 2.0) node{${\scriptstyle X_{k+2}}$};
 \draw [line width=0.22mm, dotted] (2.3,1.55) -- (2.3,1.2);
 \draw (2.3,0.8) node{${\scriptstyle X_{2k}}$};
 \draw (5.35,2.6) node{${\scriptstyle X_{2k+1}}$};
 \draw (5.35,2.0) node{${\scriptstyle X_{2k+2}}$};
 \draw [line width=0.22mm, dotted] (5.35,1.55) -- (5.35,1.2);
 \draw (5.35,0.8) node{${\scriptstyle X_{3k}}$};
 \draw (8.15,2.6) node{${\scriptstyle X_{(L-1)k+1}}$};
 \draw (8.15,2.0) node{${\scriptstyle X_{(L-1)k+2}}$};
 \draw [line width=0.22mm, dotted] (5.35,1.55) -- (5.35,1.2);
 \draw (8.15,0.8) node{${\scriptstyle X_{Lk}}$};
 \draw (11.3,2.6) node{${\scriptstyle X_{Lk+1}}$};
 \draw (11.3,2.0) node{${\scriptstyle X_{Lk+2}}$};
 \draw [line width=0.22mm, dotted] (11.3,1.55) -- (11.3,1.2);
 \draw (11.3,0.8) node{${\scriptstyle X_{(L+1)k}}$};
\draw (0.8,3.25) node{$\pi_1$};
 \draw (3.8,3.25) node{$\pi_2$};
 \draw (9.75,3.25) node{$\pi_{L}$};
 \draw (-2.3,2.4) node{$g_1$};
 \draw (-2.3,1.8) node{$g_2$};
\draw (-2.3,0.6) node{$g_k$};
 \draw (13.1,2.4) node{$g_{k+1}$};
 \draw (13.1,1.8) node{$g_{k+2}$};
 \draw (13.1,0.6) node{$g_{2k}$};
   \draw [line width=0.27mm] (-1.6,2.4) -- (0,2.4);
   \draw [line width=0.27mm] (-1.6,1.8) -- (0,1.8);
   \draw [line width=0.27mm] (-1.6,0.6) -- (0,0.6);
   \draw [line width=0.27mm] (1.5,2.4) -- (3.1,2.4);  
   \draw [line width=0.27mm] (1.5,1.8) -- (3.1,1.8);  
   \draw [line width=0.27mm] (1.5,0.6) -- (3.1,0.6);  
   \draw [line width=0.27mm] (4.6,2.4) -- (6.2,2.4);  
   \draw [line width=0.27mm] (4.6,1.8) -- (6.2,1.8);  
   \draw [line width=0.27mm] (4.6,0.6) -- (6.2,0.6);
   \draw [line width=0.27mm] (7.4,2.4) -- (9,2.4);  
   \draw [line width=0.27mm] (7.4,1.8) -- (9,1.8);  
   \draw [line width=0.27mm] (7.4,0.6) -- (9,0.6);  
   \draw [line width=0.27mm] (10.5,2.4) -- (12.1,2.4);  
   \draw [line width=0.27mm] (10.5,1.8) -- (12.1,1.8);  
   \draw [line width=0.27mm] (10.5,0.6) -- (12.1,0.6);  
   \draw [line width=0.27mm, dotted] (6.5,2.4) -- (7.1,2.4);
   \draw [line width=0.27mm, dotted] (6.5,1.8) -- (7.1,1.8);
   \draw [line width=0.27mm, dotted] (6.5,0.6) -- (7.1,0.6);
  \end{tikzpicture}
  \vspace{1.0ex}
  \caption{\label{fig:Primalfactors}
The NFG for the ladder graph with factorization in~(\ref{eqn:globalprimal}). 
The factors $\{\pi_\ell, 1 \le \ell \le L\}$ are represented by big boxes and the variables $\{X_i, 1 \le  i \le N\}$ 
are represented by edges.
The small boxes represent the constant one factors $\{g_i, 1 \le  i \le 2k\}$.}
\end{figure*}

\begin{figure*}[t]
  \centering
  \begin{tikzpicture}[scale=0.72]

%
 %
 \draw (0,2.4) node{$X_1$};
  \draw (2,2.4) node{$X_3$};
   \draw (4,2.4) node{$X_5$};
     \draw (6,2.4) node{$X_{2L-1}$};
       \draw (8,2.4) node{$X_{2L+1}$};
         \draw (0,-0.4) node{$X_2$};
           \draw (2,-0.4) node{$X_4$};
           \draw (4,-0.4) node{$X_6$};
           \draw (6,-0.4) node{$X_{2L}$};
           \draw (8,-0.4) node{$X_{2L+2}$};
\draw[fill=black] (0,0) circle (1.2mm);
\draw[fill=black] (0,2) circle (1.2mm);
\draw[fill=black] (2,0) circle (1.2mm);
\draw[fill=black] (2,2) circle (1.2mm);
   \draw [line width=0.22mm] (0,0) -- (0,2);
   \draw [line width=0.22mm] (0,2) -- (2,2);
   \draw [line width=0.22mm] (0,0) -- (2,0);  
   \draw [line width=0.22mm] (2,0) -- (2,2);
   \draw [line width=0.22mm] (0,0) -- (2,2);  
   \draw [line width=0.22mm] (2,0) -- (0,2);
\draw[fill=black] (4,0) circle (1.2mm);
\draw[fill=black] (4,2) circle (1.2mm);
   \draw [line width=0.22mm] (2,0) -- (4,0);
   \draw [line width=0.22mm] (2,2) -- (4,2);
   \draw [line width=0.22mm] (4,0) -- (4,2);  
   \draw [line width=0.22mm] (2,2) -- (4,0);
   \draw [line width=0.22mm] (2,0) -- (4,2);  
   \draw [line width=0.22mm, dotted] (4.4,0) -- (5.6,0);
   \draw [line width=0.22mm, dotted] (4.4,2) -- (5.6,2);
\draw[fill=black] (6,0) circle (1.2mm);
\draw[fill=black] (6,2) circle (1.2mm);
\draw[fill=black] (8,0) circle (1.2mm);
\draw[fill=black] (8,2) circle (1.2mm);
   \draw [line width=0.22mm] (6,0) -- (8,0);
   \draw [line width=0.22mm] (6,2) -- (8,2);
   \draw [line width=0.22mm] (6,0) -- (6,2);  
   \draw [line width=0.22mm] (8,0) -- (8,2);
   \draw [line width=0.22mm] (6,0) -- (8,2);  
   \draw [line width=0.22mm] (6,2) -- (8,0);  
  \end{tikzpicture}
  \vspace{1.0ex}
  \caption{\label{fig:Primal}
Dependencies among the variables in Fig.~\ref{fig:Primalfactors} for $k=2$, where all entries of 
the local precision matrices $\{\mathsf{\Sigma}^{-1}_\ell, 1 \le \ell \le L\}$ are non-zero.}
\end{figure*}
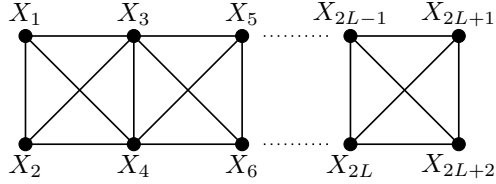

\section{The model}
\label{sec:Model}

First, we consider the following PDF defined by the product of $L$ local factors as
\begin{equation}
\label{eqn:Altglobalprimal}
p(\x) = \frac{1}{Z_f}\prod_{\ell=1}^L f_{\ell}\big(x_{(\ell-1)k+1}, x_{(\ell-1)k+2}, \ldots, x_{(\ell+1)k}\big),
\end{equation}
where $Z_f$ is the appropriate normalization constant. Each local factor 
$f_\ell\big(\x_{(\ell-1)k+1}^{(\ell+1)k}\big)$ is given by
\begin{multline}
\label{eqn:fl}
f_{\ell}\big(x_{(\ell-1)k+1},\ldots, x_{(\ell+1)k}\big)  = \\ 
\textrm{exp}\Big(-\frac{1}{2}
\big[x_{(\ell-1)k+1}, \, \ldots, \, x_{(\ell+1)k}\big] 
\,\mathsf{\Sigma}_{\ell}^{-1} \begin{bmatrix} 
x_{(\ell-1)k+1} \\ \vdots \\ x_{(\ell+1)k}
\end{bmatrix} \Big)
\end{multline}
for $1 \le \ell \le L$.

Here, $\mathsf{\Sigma}^{-1}_\ell$ is the local precision matrix associated with $f_\ell$. 
In this setup, each local factor contains $2k$ variables, and for $1 \le \ell < L$ two consecutive factors $f_{\ell}$ and $f_{\ell+1}$  
have exactly $k$ variables $\x_{\ell k+1}^{(\ell+1)k}$ in common. 
Therefore, the total number of variables in the model $N$ is given by
\begin{equation}
\label{eqn:total}
N = (L+1)k
\end{equation}

Clearly $p(\x)$ in (\ref{eqn:Altglobalprimal}) follows a zero-mean $N$-variate Gaussian distribution. Let
\begin{equation}
\label{eqn:ZF}
Z_f = \textrm{det}(2\pi \mathsf{\Sigma})^{1/2},
\end{equation}
where $\mathsf{\Sigma}$ is the covariance matrix 
associated with $p(\x)$.

Next, we write the PDF in (\ref{eqn:Altglobalprimal}) in a slightly different form as
\begin{equation}
\label{eqn:globalprimal}
\pi(\x) = \frac{1}{Z}\prod_{\ell = 1}^L \pi_{\ell}\big(x_{(\ell-1)k+1}, \ldots, x_{(\ell+1)k}\big),
\end{equation}
where $Z$ is the normalization constant, and $\pi_{\ell}\big(\x_{(\ell-1)k+1}^{(\ell+1)k}\big)$
\begin{multline}
\label{eqn:pit}
\pi_{\ell}\big(x_{(\ell-1)k+1}, \ldots, x_{(\ell+1)k}\big)  = \\
\frac{1}{Z_\ell}f_{\ell}\big(x_{(\ell-1)k+1}, \ldots, x_{(\ell+1)k}\big)
\end{multline}
has a zero-mean \mbox{2$k$-variate} Gaussian distribution for $1 \le \ell \le L$. 
Fom (\ref{eqn:fl}), the local normalization 
constant $Z_\ell$ is
\begin{equation}
\label{eqn:Zl}
Z_\ell = \textrm{det}(2\pi \mathsf{\Sigma}_\ell)^{1/2}
\end{equation}

The global factorization~(\ref{eqn:globalprimal}) creates a ladder graph as a concatenation of $L$ 
blocks (rungs), in which, each block $\pi_\ell$ has a zero-mean \mbox{2$k$-variate} Gaussian distribution.

The normalization constants $Z$ and $Z_f$ are closely related. It is straightforward to show that
\begin{align}
Z & =  \frac{Z_f}{\prod_{\ell =1}^L Z_\ell} \label{eqn:detS1} \\
   & = \frac{\textrm{det}(2\pi \mathsf{\Sigma})^{1/2}}{\prod_{\ell =1}^L\textrm{det}(2\pi \mathsf{\Sigma}_\ell)^{1/2}} \label{eqn:detS2} 
\end{align}

Our aim in this paper is to propose an efficient method to compute the exact value 
of $\textrm{det}(\mathsf{\Sigma})$, which from (\ref{eqn:detS2}), boils down to an efficient computation of $Z$. 
Although the problems of computing $Z_f$ and $Z$ essentially have the same computational complexity, it is more 
convenient to describe our approach in 
terms of $\pi(\x)$ in (\ref{eqn:globalprimal}) rather than $p(\x)$ in (\ref{eqn:Altglobalprimal}).

\section{Assumptions and Graphical Models}
\label{sec:Assume}

We suppose the following assumptions hold for the local covariance and the local precision matrices:
\begin{enumerate}[I.]
\item All local covariance matrices are $2k\times 2k$ symmetric positive-definite matrices, i.e.,
\begin{equation}
\label{eqn:podlocal}
\mathsf{\Sigma}_\ell \succ 0, \quad \forall \ell, \; 1 \le \ell \le L
\end{equation}
\item  All local precision matrices $\{\mathsf{\Sigma}^{-1}_\ell, 1 \le \ell \le L\}$ have graphical representations with cycles. 

\item All local covariance matrices $\{\mathsf{\Sigma}_\ell, 1 \le \ell \le L\}$ and their concatenation have cycle-free graphical 
representations. 
\end{enumerate}

We will use NFGs as graphical models~\cite{Forney:01,AY:2011}.
The NFG for the factorization (\ref{eqn:globalprimal}) is illustrated in Fig.~\ref{fig:Primalfactors}, in which, the 
big boxes represent the factors $\{\pi_\ell, 1 \le \ell \le L\}$ as in (\ref{eqn:pit}), the 
edges represent the variables $\{X_i, 1 \le  i \le N\}$, 
and the small boxes represent additional constant one factors $\{g_i, 1 \le  i \le 2k\}$. 
In Fig.~\ref{fig:Primalfactors}, we have attached $2k$ additional univariate constant one factors $\{g_i, 1 \le  i \le 2k\}$ to the NFG, 
$k$ such factors to the leftmost side, and the remaining
$k$ factors to the rightmost side of the ladder. 
Obviously, these extra factors do not affect the value of $Z$, 
however, including them in the model will 
facilitate our analysis in Section~\ref{sec:Trans}. 

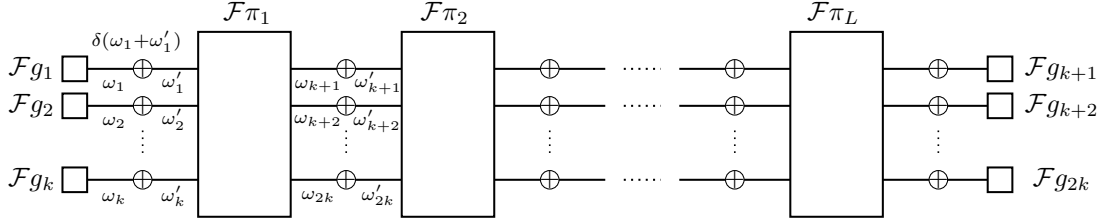
\begin{figure*}[t]
  \centering
  \begin{tikzpicture}[scale=0.82]

 \draw [line width=0.27mm] (0, 0) rectangle (1.5,3.0);
 \draw [line width=0.27mm] (3.3, 0) rectangle (4.8,3.0);
\draw [line width=0.27mm] (9.6, 0) rectangle (11.1,3.0);
\draw [line width=0.27mm] (-2.2, 2.2) rectangle (-1.8,2.6);
\draw [line width=0.27mm] (-2.2, 1.6) rectangle (-1.8,2.0);
\draw [line width=0.27mm] (-2.2, 0.4) rectangle (-1.8,0.8);
\draw [line width=0.27mm] (12.8, 2.2) rectangle (13.2,2.6);
\draw [line width=0.27mm] (12.8, 1.6) rectangle (13.2,2.0);
\draw [line width=0.27mm] (12.8, 0.4) rectangle (13.2,0.8);
 \node at (-0.9, 2.4){$+$};
 \node at (-0.9, 1.8){$+$};
 \node at (-0.9, 0.6){$+$};
 \node at (2.4, 2.4){$+$};
 \node at (2.4, 1.8){$+$};
 \node at (2.4, 0.6){$+$};
 \node at (5.7, 2.4){$+$};
 \node at (5.7, 1.8){$+$};
 \node at (5.7, 0.6){$+$};
 \node at (8.7, 2.4){$+$};
 \node at (8.7, 1.8){$+$};
 \node at (8.7, 0.6){$+$};
 \node at (12, 2.4){$+$};
 \node at (12, 1.8){$+$};
 \node at (12, 0.6){$+$};
 \draw (-1.36,2.13) node{${\scriptstyle \omega_1}$};
 \draw (-0.4,2.18) node{${\scriptstyle \omega'_1}$};
\draw (-1.36,1.53) node{${\scriptstyle \omega_2}$};
 \draw (-0.4,1.58) node{${\scriptstyle \omega'_2}$};
\draw (-1.36,0.3) node{${\scriptstyle \omega_k}$};
 \draw (-0.4,0.34) node{${\scriptstyle \omega'_k}$};
\draw (1.95,2.12) node{${\scriptstyle \omega_{k+1}}$};
 \draw (2.92,2.17) node{${\scriptstyle \omega'_{k+1}}$};
\draw (1.94,1.55) node{${\scriptstyle \omega_{k+2}}$};
 \draw (2.90,1.56) node{${\scriptstyle \omega'_{k+2}}$};
\draw (1.95,0.33) node{${\scriptstyle \omega_{2k}}$};
 \draw (2.92,0.36) node{${\scriptstyle \omega'_{2k}}$};
\draw (0.8,3.3) node{$\mathcal{F} \pi_1$};
 \draw (4.0,3.3) node{$\mathcal{F} \pi_2$};
 \draw (10.3,3.3) node{$\mathcal{F} \pi_{L}$};
\draw(-1.0, 2.85) node{${\scriptstyle \delta(\omega_1+\omega'_1)}$};
 \draw (-2.7,2.4) node{$\mathcal{F} g_1$};
 \draw (-2.7,1.8) node{$\mathcal{F} g_2$};
\draw (-2.7,0.6) node{$\mathcal{F} g_k$};
 \draw (14,2.4) node{$\mathcal{F} g_{k+1}$};
\draw (14,1.8) node{$\mathcal{F} g_{k+2}$};
\draw (14,0.6) node{$\mathcal{F} g_{2k}$};
   \draw [line width=0.27mm] (-1.8,2.4) -- (-1.05,2.4);  
   \draw [line width=0.27mm] (-0.75,2.4) -- (0,2.4); 
   \draw (-0.9,2.4) circle (0.15cm); 
   \draw [line width=0.27mm] (-1.8,1.8) -- (-1.05,1.8);
   \draw [line width=0.27mm] (-0.75,1.8) -- (0,1.8);
   \draw (-0.9,1.8) circle (0.15cm); 
   \draw [line width=0.22mm, dotted] (-0.9,1.4) -- (-0.9,1.0);
   \draw [line width=0.27mm] (-1.8,0.6) -- (-1.05,0.6);  
   \draw [line width=0.27mm] (-0.75,0.6) -- (0,0.6);  
   \draw (-0.9,0.6) circle (0.15cm); 
   \draw [line width=0.27mm] (1.5,2.4) -- (2.25,2.4);  
   \draw [line width=0.27mm] (2.55,2.4) -- (3.3,2.4);    
   \draw (2.4,2.4) circle (0.15cm);  
  \draw [line width=0.27mm] (1.5,1.8) -- (2.25,1.8);  
   \draw [line width=0.27mm] (2.55,1.8) -- (3.3,1.8);    
   \draw (2.4,1.8) circle (0.15cm);  
  \draw [line width=0.27mm] (1.5,0.6) -- (2.25,0.6);  
   \draw [line width=0.27mm] (2.55,0.6) -- (3.3,0.6);    
   \draw (2.4,0.6) circle (0.15cm);  
   \draw [line width=0.22mm, dotted] (2.4,1.4) -- (2.4,1.0);
   \draw [line width=0.27mm] (4.8,2.4) -- (5.55,2.4);  
   \draw [line width=0.27mm] (5.85,2.4) -- (6.6,2.4);    
   \draw (5.7,2.4) circle (0.15cm);  
  \draw [line width=0.27mm] (4.8,1.8) -- (5.55,1.8);  
   \draw [line width=0.27mm] (5.85,1.8) -- (6.6,1.8);    
  \draw (5.7,1.8) circle (0.15cm);  
  \draw [line width=0.27mm] (4.8,0.6) -- (5.55,0.6);  
  \draw [line width=0.27mm] (5.85,0.6) -- (6.6,0.6);    
\draw (5.7,0.6) circle (0.15cm);  
   \draw [line width=0.22mm, dotted] (5.7,1.4) -- (5.7,1.0);
   \draw [line width=0.27mm] (7.8,2.4) -- (8.55,2.4);  
   \draw [line width=0.27mm] (8.85,2.4) -- (9.6,2.4);    
   \draw (8.7,2.4) circle (0.15cm);  
  \draw [line width=0.27mm] (7.8,1.8) -- (8.55,1.8);  
   \draw [line width=0.27mm] (8.85,1.8) -- (9.6,1.8);    
  \draw (8.7,1.8) circle (0.15cm);  
  \draw [line width=0.27mm] (7.8,0.6) -- (8.55,0.6);  
  \draw [line width=0.27mm] (8.85,0.6) -- (9.6,0.6);    
\draw (8.7,0.6) circle (0.15cm);  
   \draw [line width=0.22mm, dotted] (8.7,1.4) -- (8.7,1.0);
  \draw [line width=0.27mm] (11.1,2.4) -- (11.85,2.4);  
   \draw [line width=0.27mm] (12.15,2.4) -- (12.8,2.4);    
   \draw (12,2.4) circle (0.15cm);  
  \draw [line width=0.27mm] (11.1,1.8) -- (11.85,1.8);  
   \draw [line width=0.27mm] (12.15,1.8) -- (12.8,1.8);    
   \draw (12,1.8) circle (0.15cm);  
  \draw [line width=0.27mm] (11.1,0.6) -- (11.85,0.6);  
\draw [line width=0.27mm] (12.15,0.6) -- (12.8,0.6);    
\draw (12,0.6) circle (0.15cm);  
 \draw [line width=0.22mm, dotted] (12,1.4) -- (12,1.0);
   \draw [line width=0.27mm, dotted] (6.9,2.4) -- (7.5,2.4);
   \draw [line width=0.27mm, dotted] (6.9,1.8) -- (7.5,1.8);
   \draw [line width=0.27mm, dotted] (6.9,0.6) -- (7.5,0.6);
  \end{tikzpicture}
  \vspace{1.0ex}
  \caption{\label{fig:Dualfactors}
The dual NFG of Fig.~\ref{fig:Primalfactors}. 
The big boxes represent the factors $\{\mathcal{F}\pi_\ell, 1 \le \ell \le L\}$ in (\ref{eqn:gausstrans}), the edges 
represent the variables $\{\bigomega_i,\bigomega'_i, 1 \le  i \le N\}$, and the small boxes represent the factors 
$\{\mathcal{F} g_i, 1 \le  i \le 2k\}$ given by (\ref{eqn:deltatrans}). The factors denoted by $\oplus$ transform each edge to a sign-inverting edge.}
\end{figure*}
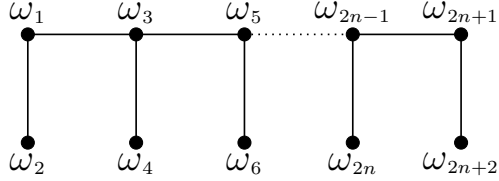
\begin{figure*}[t]
  \centering
  \begin{tikzpicture}[scale=0.72]

 \draw (0,2.4) node{$\bigomega_1$};
  \draw (2,2.4) node{$\bigomega_3$};
   \draw (4,2.4) node{$\bigomega_5$};
     \draw (6,2.4) node{$\bigomega_{2n-1}$};
       \draw (8,2.4) node{$\bigomega_{2n+1}$};
         \draw (0,-0.4) node{$\bigomega_2$};
           \draw (2,-0.4) node{$\bigomega_4$};
           \draw (4,-0.4) node{$\bigomega_6$};
           \draw (6,-0.4) node{$\bigomega_{2n}$};
           \draw (8,-0.4) node{$\bigomega_{2n+2}$};
\draw[fill=black] (0,0) circle (1.2mm);
\draw[fill=black] (0,2) circle (1.2mm);
\draw[fill=black] (2,0) circle (1.2mm);
\draw[fill=black] (2,2) circle (1.2mm);
   \draw [line width=0.22mm] (0,0) -- (0,2);
   \draw [line width=0.22mm] (0,2) -- (2,2);
   \draw [line width=0.22mm] (2,0) -- (2,2);
\draw[fill=black] (4,0) circle (1.2mm);
\draw[fill=black] (4,2) circle (1.2mm);
   \draw [line width=0.22mm] (2,2) -- (4,2);
   \draw [line width=0.22mm] (4,0) -- (4,2);  
   \draw [line width=0.22mm, dotted] (4,2) -- (6,2);
\draw[fill=black] (6,0) circle (1.2mm);
\draw[fill=black] (6,2) circle (1.2mm);
\draw[fill=black] (8,0) circle (1.2mm);
\draw[fill=black] (8,2) circle (1.2mm);
   \draw [line width=0.22mm] (6,2) -- (8,2);
   \draw [line width=0.22mm] (6,0) -- (6,2);  
   \draw [line width=0.22mm] (8,0) -- (8,2);
  \end{tikzpicture}
  \vspace{1.0ex}
  \caption{\label{fig:DualCompact}
Dependencies among the variables in Fig.~\ref{fig:Dualfactors} for $k = 2$. The local covariance matrices have cycle-free graphical 
representations with structure as in~(\ref{eqn:covariance}).}
\end{figure*}

Alternatively, we can illustrate the dependencies among the variables in Fig.~\ref{fig:Primalfactors} with a 
more compact graphical representation. 
By Assumption II, the local precision matrices 
\begin{equation}
\{\mathsf{\Sigma}^{-1}_\ell, 1 \le \ell \le L\}
\end{equation}
have graphical model representations that contain cycles. Let $k=2$, and assume that all entries of the $4\times 4$ local 
precision matrices 
are non-zero. Each block (rung) will therefore have a fully-connected graphical representation (as every pair of distinct variables is 
connected by an edge). Concatenation of these blocks will then form the ladder graph depicted in Fig.~\ref{fig:Primal}.


It is clear from Figs.~\ref{fig:Primalfactors} and~\ref{fig:Primal} that, in general, the graphical model 
of (\ref{eqn:globalprimal}) may contain many 
short cycles. Next, we will consider the dual NFG of the Gaussian ladder graph in Fig.~\ref{fig:Primalfactors}.

\section{The Cycle-Free Dual NFG}
\label{sec:Trans}

Following~\cite{Forney:18}, we can construct the dual of the NFG in Fig.~\ref{fig:Primalfactors}
by employing these three steps

\begin{itemize}
\item Replace each variable $X_\ell$ by the dual variable $\bigomega_\ell$.
\item Replace each factor by its Fourier transform.
\item Replace each edge by a sign-inverting edge.
\end{itemize}

The Fourier transform of a function $f(\x) \colon \R^n \rightarrow \C$ is the function
$(\mathcal{F}f)(\pmb{\omega}) \colon \R^n \rightarrow \C$ given by
\begin{equation}
\label{eqn:ft}
(\mathcal{F}f)(\pmb{\omega}) = \int_{-\infty}^{\infty} f(\x)e^{-\mathrm{i}\,\pmb{\omega}^\intercal \x}d\x,
\end{equation}
where $\C$ is the set of complex numbers, $\mathrm{i} = \sqrt{-1}$ denotes the unit imaginary number, and 
$\pmb{\omega} = (\omega_1, \omega_2, \ldots, \omega_n)^\intercal$.

The Fourier transform of the constant one factors
\begin{equation} 
\{g_i, 1 \le i \le 2k\}
\end{equation}
are Dirac delta functions up to scale. 
Indeed 
\begin{equation}
\label{eqn:deltatrans}
(\mathcal{F}g_1)(\omega_1) = 2\pi\delta(\omega_1), 
\end{equation}
and similarly for $\mathcal{F}g_2, \mathcal{F}g_3, \ldots, \mathcal{F}g_{2k}$.

The factors $\mathcal{F}g_1,\mathcal{F}g_2, \ldots, \mathcal{F}g_{2k}$ 
will replace the factors $g_1, g_2, \ldots, g_{2k}$ in the dual NFG. Again, $k$ of these factors are 
connected to the leftmost side, and 
the remaining $k$ factors to the rightmost side of the dual graph.  

The Fourier transform of $\pi_\ell$ in~(\ref{eqn:pit}) is 
\begin{multline} 
\label{eqn:gausstrans}
(\mathcal{F}\pi_\ell)\big(\omega_{(\ell-1)k+1}, \ldots, \omega_{(\ell+1)k}\big) = \\
\textrm{exp}\bigg(-\frac{1}{2}
\big[\omega_{(\ell-1)k+1},\ldots, \, \omega_{(\ell+1)k}\big] 
\,\mathsf{\Sigma}_\ell \begin{bmatrix} 
\omega_{(\ell-1)k+1} \\ \vdots \\ \omega_{(\ell+1)k}
\end{bmatrix}  \bigg)
\end{multline}

Here $\mathsf{\Sigma}_\ell$ is the covariance matrix associated with $\pi_\ell$. 
Each factor $\pi_\ell$ is then replaced 
by $\mathcal{F}\pi_\ell$ in the dual NFG.
The crucial property of the Fourier transform of the multivariate Gaussian distribution is that 
in $\mathcal{F}\pi_\ell$ the covariance matrix is replaced by
the precision matrix in the exponent.\footnote{Since $\pi_\ell$ follows a zero-mean 
multivariate Gaussian distribution, its Fourier transform $\mathcal{F}\pi_\ell$ is identical to its characteristic function~\cite{anderson1958introduction}.}

Sign-inverting edges are created by inserting a $\oplus-$factor in the middle of each edge. 
The $\oplus-$factors will impose the constraint that the addition of their two arguments should be zero in $\R$. 
The multiplication of all the $\oplus-$factors can be expressed by
\begin{equation}
\prod_{i=1}^{N}\delta(\omega_i + \omega'_i),
\end{equation}
where $N$ is the total number of variables as in (\ref{eqn:total}).

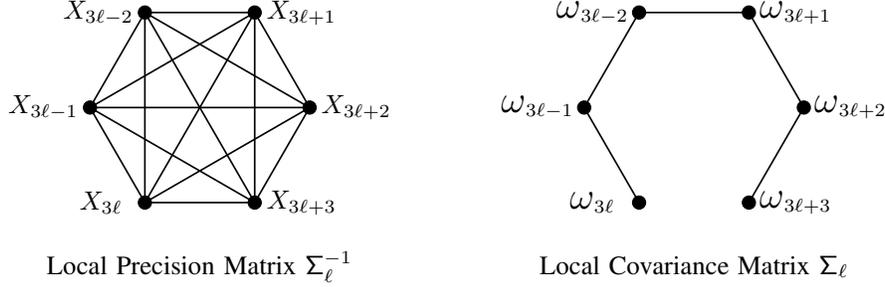
\begin{figure*}[t]
  \centering
  \begin{tikzpicture}[scale=0.73]

\draw[fill=black] (2,0) circle (1.2mm);
\draw[fill=black] (-2,0) circle (1.2mm);
\draw[fill=black] (1,1.732) circle (1.2mm);
\draw[fill=black] (-1,1.732) circle (1.2mm);
\draw[fill=black] (-1,-1.732) circle (1.2mm);
\draw[fill=black] (1,-1.732) circle (1.2mm);
\draw [line width=0.22mm] (2,0) -- (-2,0);
\draw [line width=0.22mm] (2,0) -- (1,1.732);
\draw [line width=0.22mm] (2,0) -- (-1,1.732);
\draw [line width=0.22mm] (2,0) -- (-1,-1.732);
\draw [line width=0.22mm] (2,0) -- (1,-1.732);
\draw [line width=0.22mm] (-2,0) -- (1,1.732);
\draw [line width=0.22mm] (-2,0) -- (-1,1.732);
\draw [line width=0.22mm] (-2,0) -- (-1,-1.732);
\draw [line width=0.22mm] (-2,0) -- (1,-1.732);
\draw [line width=0.22mm] (1,1.732) -- (-1,1.732);
\draw [line width=0.22mm] (1,1.732) -- (-1,-1.732);
\draw [line width=0.22mm] (1,1.732) -- (1,-1.732);
\draw [line width=0.22mm] (-1,1.732) -- (-1,-1.732);
\draw [line width=0.22mm] (-1,1.732) -- (1,-1.732);
\draw [line width=0.22mm] (1,-1.732) -- (-1,-1.732);
 \draw (2.85,0) node{$X_{3\ell+2}$};
 \draw (-2.85,0) node{$X_{3\ell-1}$};
\draw (1.85,1.73) node{$X_{3\ell+1}$};
 \draw (-1.85,1.73) node{$X_{3\ell-2}$};
\draw (1.85,-1.73) node{$X_{3\ell+3}$};
 \draw (-1.8,-1.73) node{$X_{3\ell}$};
\draw[fill=black] (11,0) circle (1.2mm);
\draw[fill=black] (7,0) circle (1.2mm);
\draw[fill=black] (10,1.732) circle (1.2mm);
\draw[fill=black] (8,1.732) circle (1.2mm);
\draw[fill=black] (8,-1.732) circle (1.2mm);
\draw[fill=black] (10,-1.732) circle (1.2mm);
\draw [line width=0.22mm] (10,-1.732) -- (11,0);
\draw [line width=0.22mm] (11,0) -- (10,1.732);
\draw [line width=0.22mm] (10,1.732) -- (8,1.732);
\draw [line width=0.22mm] (8,1.732) -- (7,0);
\draw [line width=0.22mm] (7,0) -- (8,-1.732);
 \draw (0,-2.9) node{Local Precision Matrix $\mathsf{\Sigma}^{-1}_\ell$};
 \draw (9,-2.9) node{Local Covariance Matrix $\mathsf{\Sigma}_\ell$};
 \draw (11.85,0) node{$\bigomega_{3\ell+2}$};
 \draw (6.15,0) node{$\bigomega_{3\ell-1}$};
\draw (10.85,1.73) node{$\bigomega_{3\ell+1}$};
 \draw (7.15,1.73) node{$\bigomega_{3\ell-2}$};
\draw (10.85,-1.73) node{$\bigomega_{3\ell+3}$};
 \draw (7.15,-1.73) node{$\bigomega_{3\ell}$};
%
%

  \end{tikzpicture}
  \vspace{1.0ex}
  \caption{\label{fig:Transf2}
The contrast between the dependency among the variables associated with the local precision matrix
$\mathsf{\Sigma}^{-1}_\ell$ in the primal domain (left) and among the variables associated with the 
local covariance matrix
$\mathsf{\Sigma}_\ell$ in the dual domain (right), for $k=3$.}
\end{figure*}

The dual of the NFG in Fig.~\ref{fig:Primalfactors} is illustrated in Fig.~\ref{fig:Dualfactors}, where the 
big boxes represent $\{\mathcal{F}\pi_\ell, 1 \le \ell \le L\}$ as in~(\ref{eqn:gausstrans}), the edges denote
the variables $\{\bigomega_i,\bigomega'_i, 1 \le  i \le N\}$, the small boxes correspond to the additional 
factors $\{\mathcal{F}g_i, 1 \le i \le 2k\}$ given by (\ref{eqn:deltatrans}), and the $\oplus-$factors 
in the middle of each edge denote sign-inverters. 

The global PDF in the dual domain $\pi'$ is given by 
\begin{multline}
\label{eqn:globaldual}
\pi'(\pmb{\omega}) = \\
\frac{1}{Z'}\prod_{i=1}^{N}\delta(\omega_i + \omega'_i)
\prod_{i=1}^k (\mathcal{F}g_i)(\omega_i)\prod_{i=1}^{k} (\mathcal{F}g_{k+i})(\omega'_{Lk+i})
\cdot\\ 
\prod_{\ell = 1}^L (\mathcal{F}\pi_\ell)\big(\omega'_{(\ell-1)k+1}, \ldots, 
\omega'_{\ell k}, \omega_{\ell k +1}, \ldots, \omega_{(\ell+1)k}\big),
\end{multline}
where $Z'$ is the appropriate normalization constant.

In the dual domain, we can again show the dependencies among the variables  
using a more compact graphical representation.
By Assumption III in Section~\ref{sec:Trans}, the local covariance 
matrices $\{\mathsf{\Sigma}_\ell, 1 \le \ell \le L\}$ 
have cycle-free graphical representations. Set $k=2$, and assume that all the local covariance matrices have the following structure
\begin{equation}
\label{eqn:covariance}
\mathsf{\Sigma}_\ell = 
\begin{bmatrix}
\ast & \ast & \ast & 0 \\
\ast & \ast & 0 & 0 \\
\ast & 0 & \ast & \ast \\
0 & 0 & \ast & \ast
\end{bmatrix}
\end{equation}
where $\ast$ denotes a non-zero real entry. In the corresponding graphical model, each block (rung) and their 
concatenation are therefore 
cycle-free -- as illustrated in~Fig.~\ref{fig:DualCompact}.

The factors $\{\mathcal{F}g_i, 1 \le i \le 2k\}$ set the value of $2k$ out of $N$ independent variables 
in (\ref{eqn:globaldual}) 
to zero. Therefore, $\pi'$ 
follows a zero-mean $(N-2k)$-variate Gaussian distribution. 
Let us denote the covariance matrix 
associated with $\pi'(\x)$ by $\mathsf{\Sigma'}$. Furthermore, the factors $\{\mathcal{F}g_i, 1 \le i \le 2k\}$
contribute a multiplicative factor of $(2\pi)^{2k}$ to~(\ref{eqn:globaldual}).

It follows that
\begin{align}
Z' & = (2\pi)^{2k} \textrm{det}(2\pi \mathsf{\Sigma'})^{1/2} \label{eqn:Zdual1} \\
   & = (2\pi)^{k+N/2} \textrm{det}(\mathsf{\Sigma'})^{1/2} \label{eqn:Zdual2}
\end{align}

According to the NFG duality theorem, the normalization constants $Z'$ and $Z$ are equal up to scale.
Indeed
\begin{equation}
\label{eqn:FGduality}
Z' = (2\pi)^{N}Z
\end{equation}
For more details, see~\cite{AY:2011, Forney:18}.

\begin{proposition} \label{prop:DetRelations}
The $\textrm{det}(\mathsf{\Sigma})$ can be expressed in terms of $\textrm{det}(\mathsf{\Sigma'})$ 
and $\{\textrm{det}(\mathsf{\Sigma}_\ell), 1 \le \ell \le L\}$ as
\begin{equation}
\textrm{det}(\mathsf{\Sigma}) = \textrm{det}(\mathsf{\Sigma'})\prod_{\ell =1}^L\textrm{det}(\mathsf{\Sigma}_\ell)
\label{eqn:DetTransformation}
\end{equation}
\end{proposition}

\begin{trivlist}
\item \emph{Proof. } The local covariance matrices $\{\mathsf{\Sigma}_\ell, 1 \le \ell \le L\}$ are of size $2k\times 2k$ and 
$\mathsf{\Sigma}$ is of size $N\times N$. 
Thus, from (\ref{eqn:total}), (\ref{eqn:detS2}), and (\ref{eqn:FGduality}) we obtain
\begin{align}
Z' & = (2\pi)^{N}\frac{\textrm{det}(2\pi \mathsf{\Sigma})^{1/2}}{\prod_{\ell =1}^L\textrm{det}(2\pi \mathsf{\Sigma}_\ell)^{1/2}} \\
   & = (2\pi)^{k + N/2}\frac{\textrm{det}(\mathsf{\Sigma})^{1/2}}{\prod_{\ell =1}^L\textrm{det}(\mathsf{\Sigma}_\ell)^{1/2}}
\label{eqn:ProofProp}
\end{align}
Combining (\ref{eqn:Zdual2}) and (\ref{eqn:ProofProp}) completes the proof.
\hfill$\blacksquare$
\end{trivlist}

It is therefore possible to compute $Z'$ and $\textrm{det}(\mathsf{\Sigma'})$ exactly (e.g., via the Gaussian belief 
propagation algorithm~\cite{weiss2001correctness}) in the cycle-free NFG as in Fig.~\ref{fig:Dualfactors}.
After this step, we can compute $\textrm{det}(\mathsf{\Sigma})$ 
from~(\ref{eqn:DetTransformation}).

We conclude that computing the exact value of $\textrm{det}(\mathsf{\Sigma})$ in ladder graphs is 
straightforward in the dual factor graph
if all local covariance matrices are ``known" to have cycle-free NFGs. 
On the contrary, if the local precision matrices have cycle-free graphical representations, 
computing $\textrm{det}(\mathsf{\Sigma})$ is easy in the primal factor graph. 
A special case of interest
is when the local precision and covariance matrices are diagonal. In this case, computing $\textrm{det}(\mathsf{\Sigma})$ is easy in both domains.


For $k=3$, the contrast between the dependency among the variables 
in Figs.~\ref{fig:Primal} and~\ref{fig:DualCompact} is depicted in Fig.~\ref{fig:Transf2}. By Assumption II, Fig.~\ref{fig:Transf2} (left) 
has cycles, while, by Assumption III, Fig.~\ref{fig:Transf2} (right) is singly-connected.

\section{Numerical Examples}
\label{sec:num}

In the first toy example, we set $k = 2$ and $L = 3$. 
For $1 \le \ell \le 3$, let 
\begin{equation}
\label{eqn:example}
\mathsf{\Sigma}_\ell = 
\begin{bmatrix}
2 & 1 & 1 & 0 \\
1 & 2 & 0 & 0 \\
1 & 0 & 2 & 1 \\
0 & 0 & 1 & 2
\end{bmatrix}
\end{equation}
where
\begin{equation}
\label{eqn:example1Precision}
\mathsf{\Sigma}^{-1}_\ell = 
\begin{bmatrix}
1.2 & -0.6 & -0.8 & 0.4 \\
-0.6 & 0.8 & 0.4 & -0.2 \\
-0.8 & 0.4 & 1.2 & -0.6 \\
0.4 & -0.2 & -0.6 & 0.8
\end{bmatrix}
\end{equation}
with $ \textrm{det}(\mathsf{\Sigma}_\ell) = 5$. 
Here $\mathsf{\Sigma}_\ell$ and $\mathsf{\Sigma}^{-1}_\ell$ satisfy all the required assumptions discussed 
in Section~\ref{sec:Assume}. 
This toy example consists of three blocks, where
each block has a four-variate Gaussian distribution with a fully-connected NFG.

From (\ref{eqn:globaldual}) 
the global PDF in the dual domain $\pi'(\pmb{\omega})$ is given by
\begin{multline*}
\pi'(\pmb{\omega}) = \\\frac{1}{Z'}
\mathcal{F}g_1(\omega_1)\mathcal{F}g_2(\omega_2)
\mathcal{F}g_3(\omega'_7)\mathcal{F}g_4(\omega'_8) \prod_{i=1}^{8}\delta(\omega_i + \omega'_i)\cdot \\ 
\mathcal{F}\pi_1(\omega'_{1}, \omega'_{2}, \omega_{3}, \omega_{4})\mathcal{F}\pi_2(\omega'_{3}, \omega'_{4}, \omega_{5}, \omega_{6})
\mathcal{F}\pi_3(\omega'_{5}, \omega'_{6}, \omega_{7}, \omega_{8}) 
\end{multline*}

Thus
\begin{multline*}
Z'  = (2\pi)^4\cdot \\\int_{\pmb{\omega}} 
\delta(\omega_1)\delta(\omega_2)\delta(\omega'_7)\delta(\omega'_8)\prod_{i=1}^{8}\delta(\omega_i + \omega'_i)\cdot
\mathcal{F}\pi_1\mathcal{F}\pi_2
\mathcal{F}\pi_3 d\pmb{\omega}
\end{multline*}
which gives
\begin{multline*}
Z'  = (2\pi)^4\int_{\omega_3, \omega_4, \omega_5, \omega_6} 
\mathcal{F}\pi_1(0, 0, \omega_{3}, \omega_{4})\\
\mathcal{F}\pi_2(-\omega_{3}, -\omega_{4}, \omega_{5}, \omega_{6})
\mathcal{F}\pi_3(-\omega_{5}, -\omega_{6}, 0, 0)d\omega_3d\omega_4d\omega_5d\omega_6
\end{multline*}
which after substituting $\mathsf{\Sigma}$ from (\ref{eqn:example}) yields
\begin{multline*}
\label{eqn:toy234}
Z' = (2\pi)^4\cdot \\ 
\int_{\omega_3, \omega_4, \omega_5, \omega_6}\!\!\!\!\!\!\!\!\!\!\!\!\textrm{exp}\Big(-\frac{1}{2}
\big[\omega_3, \omega_4, \omega_5, \omega_6\big] 
\,\mathsf{\Sigma'}^{-1} \begin{bmatrix} 
\omega_3 \\ \omega_4 \\ \omega_5 \\ \omega_6
\end{bmatrix}\Big) d\omega_3d\omega_4d\omega_5d\omega_6
\end{multline*}
with 
\begin{equation}
\mathsf{\Sigma'}^{-1} = 
\begin{bmatrix}
4 & 2 & -1 & 0 \\
2 & 4 & 0 & 0 \\
-1 & 0 & 4 & 2 \\
0 & 0 & 2 & 4
\end{bmatrix}
\end{equation}

To compute $\textrm{det}(\mathsf{\Sigma'})$, we can either apply the Gaussian belief propagation 
algorithm, or, in this toy example, directly 
compute 
\begin{equation}
\label{eqn:toy24}
\textrm{det}(\mathsf{\Sigma'}) = 1/128
\end{equation}
From (\ref{eqn:DetTransformation}) we have
\begin{equation}
\label{eqn:toy25}
\textrm{det}(\mathsf{\Sigma}) = 125/128
\end{equation}
where $\mathsf{\Sigma}$ is the $8\times 8$ covariance matrix associated with the
eight-variate Gaussian distribution in the primal NFG.

In our second example, we set $k = 3$ and $L = 2$. Let
\begin{equation}
\label{eqn:example2}
\mathsf{\Sigma}_1 = 
\begin{bmatrix}
3 & 1 & 0 & 2 & 0 & 0 \\
1 & 2 & 1 & 0 & 0 & 0 \\
0 & 1 & 3 & 0 & 0 & 0 \\
2 & 0 & 0 & 3 & 1 & 0 \\
0 & 0 & 0 & 1 & 2 & 1 \\
0 & 0 & 0 & 0 & 1 & 3 
\end{bmatrix}
\end{equation}
in the first block, where
\begin{equation}
\mathsf{\Sigma}^{-1}_1 = \frac{1}{22}
\begin{bmatrix}
30 & -18 & 6 & -25 & 15 & -5 \\
-18 & 24 & -8 & 15 & -9 & 3 \\
6 & -8 & 10 & -5 & 3 & -1 \\
-25 & 15 & -5 & 30 & -18 & 6 \\
15 & -9 & 3 & -18 & 24 & -8 \\
-5 & 3 & -1 & 6 & -8 & 10
\end{bmatrix}
\end{equation}
and $ \textrm{det}(\mathsf{\Sigma}_1) = 44$. In the second block, let
\begin{equation}
\label{eqn:example3}
\mathsf{\Sigma}_2 = 
\begin{bmatrix}
4 & 1 & 0 & -2 & 0 & 0 \\
1 & 3 & -1 & 0 & 0 & 0 \\
0 & -1 & 1 & 0 & 0 & 0 \\
-2 & 0 & 0 & 4 & 1 & 0 \\
0 & 0 & 0 & 1 & 3 & -1 \\
0 & 0 & 0 & 0 & -1 & 1 
\end{bmatrix}
\end{equation}
where
\begin{equation}
\mathsf{\Sigma}^{-1}_2 = \frac{1}{33}
\begin{bmatrix}
14 & -7 & -7 & 8 & -4 & -4 \\
-7 & 20 & 20 & -4 & 2 & 2 \\
-7 & 20 & 53 & -4 & 2 & 2 \\
8 & -4 & -4 & 14 & -7 & -7 \\
-4 & 2 & 2 & -7 & 20 & 20 \\
-4 & 2 & 2 & -7 & 20 & 53
\end{bmatrix}
\end{equation}
and $\textrm{det}(\mathsf{\Sigma}_2) = 33$. 
The covariance matrix $\mathsf{\Sigma}$ is $9\times 9$, and 
the model consists of two blocks, where
each block has a six-variate Gaussian distribution
with a fully-connected NFG.


The global PDF in the dual domain is given by 
\begin{multline*}
\pi'(\pmb{\omega}) = \frac{1}{Z'}\prod_{i=1}^{9}\delta(\omega_i + \omega'_i)\cdot \\
\mathcal{F}g_1(\omega_1)\mathcal{F}g_2(\omega_2)\mathcal{F}g_3(\omega_3)
\mathcal{F}g_4(\omega'_7)\mathcal{F}g_5(\omega'_8)\mathcal{F}g_6(\omega'_9)\cdot \\ 
\mathcal{F}\pi_1(\omega'_{1}, \omega'_{2}, \omega'_{3}, \omega_{4}, \omega_{5}, \omega_{6})
\mathcal{F}\pi_2(\omega'_{4}, \omega'_{5}, \omega'_{6}, \omega_{7}, \omega_{8}, \omega_{9})
\end{multline*}
After a little manipulation
\begin{multline*}
\label{eqn:toy32}
Z' = (2\pi)^6\cdot\\
\int_{\omega_{4}, \omega_{5}, \omega_{6}} \!\!\!\!\textrm{exp}\Big(-\frac{1}{2}\big(7\omega_4^2 + 
5\omega_5^2 + 4\omega_6^2 + 4\omega_4\omega_5\big)\Big) d\omega_4d\omega_5d\omega_6
\end{multline*}

It is then easy to compute $\textrm{det}(\mathsf{\Sigma'}) = 1/132$. Therefore, 
from (\ref{eqn:DetTransformation}) we obtain 
\begin{equation}
\label{eqn:toy35}
\textrm{det}(\mathsf{\Sigma}) = \frac{44\times 33}{132} = 11
\end{equation}

In general, the cycle-free Gaussian NFG of $\pi'(\pmb{\omega})$ in~(\ref{eqn:globaldual}) consists of 
unary factors (attached to the vertices) and pairwise 
factors (placed on the edges). 
For larger values of $N$, we can employ the Gaussian belief propagation algorithm~\cite{weiss2001correctness} to 
compute the exact value of $\textrm{det}(\mathsf{\Sigma'})$. 

\section{Future Work}

Some directions for future work include: i) applying the method to the multivariate Cauchy distribution (or other multivariate 
distributions) defined on the ladder graph, with assumptions made on the 
local dispersion matrices, ii) extending the results and using the Gaussian belief propagation in the dual 
of Gaussian Markov random fields. According to our assumptions 
in Section~\ref{sec:Assume}, 
each building block in the dual NFG of the model is cycle-free, however, arranging these 
cycle-free blocks horizontally and vertically 
will create an NFG with cycles, and 
iii) Comparing the method with existing algorithms on computing the determinant of block-tridiagonal 
matrices (see, e.g.,~\cite{meurant1992review, Molinari:08}).
Note that the precision matrices associated with ladder graphs are block-tridiagonal by construction.

\section{Conclusion}

We showed an application of the NFG duality 
theorem to continuous graphical models.
A method was proposed to solve exactly the Gaussian graphical models defined on the ladder graph
if the local covariance matrices and their concatenation have cycle-free graphical 
representations. 


\IEEEtriggeratref{20}



\end{document}